\documentclass[prl,showpacs,preprintnumbers,amsmath,amssymb,tightenlines,twocolumn,superscriptaddress]{revtex4-2}
\usepackage{graphicx}
\usepackage{dcolumn}
\usepackage{bm}
\usepackage{epsfig}
\usepackage{amsmath}
\usepackage{xcolor}
\definecolor{mycol}{RGB}{10,55,130}
\usepackage[colorlinks, linkcolor=mycol, citecolor=mycol, urlcolor=mycol, breaklinks]{hyperref}

\newcommand{\ssection}[1]{{\noi  \it #1:}}
\newcommand{\bra}[1]{\langle\,{#1}\, |}
\newcommand{\ket}[1]{|\,{#1}\,\rangle}
\newcommand{\braket}[2]{\mbox{$\langle\,{#1}\, | \,{#2}\,\rangle$}}

\newcommand{\sub}[2]{{#1}_{\mbox{\!\! \scriptsize #2}}}

\def\noi{\noindent}
\def\beq{\begin{equation}}
	\def\eeq{\end{equation}}

\newcommand{\rref}[1]{ref.~\cite{#1}}
\newcommand{\fref}[1]{Fig.~\ref{#1}}
\newcommand{\frefp}[2]{Fig.~\ref{#1}~(#2)}

\newcommand{\eref}[1]{Eq.~(\ref{#1})}

\newcommand{\cref}[1]{chapter~\ref{#1}}

\newcommand{\Cref}[1]{Chapter~\ref{#1}}

\newcommand{\beginsupplement}{%
	\setcounter{table}{0}
	\renewcommand{\thetable}{A\arabic{table}}%
	\setcounter{figure}{0}
	\renewcommand{\thefigure}{A\arabic{figure}}%
	\setcounter{equation}{0}
	\def\theequation{A\arabic{equation}}
}

\usepackage{ulem}  
\usepackage{color}

\begin{document}
	\title{Automated quantum system modeling with machine learning}
\author{K.~Mukherjee}
\affiliation{Department of Physics, Indian Institute of Science Education and Research, Bhopal, Madhya Pradesh 462 066, India}
	\affiliation{Homer L. Dodge Department of Physics and Astronomy, The University of Oklahoma, Norman, Oklahoma 73019, USA}
\affiliation{Center for Quantum Research and Technology, The University of Oklahoma, Norman, Oklahoma 73019, USA}
\author{J.~Schachenmayer}
\affiliation{CESQ/ISIS (UMR 7006), University of Strasbourg and CNRS, 67000 Strasbourg, France}
\author{S.~Whitlock}
\affiliation{ISIS (UMR 7006), University of Strasbourg and CNRS, 67000 Strasbourg, France}
\author{S.~W\"uster}
\affiliation{Department of Physics, Indian Institute of Science Education and Research, Bhopal, Madhya Pradesh 462 066, India}
\email{sebastian@iiserb.ac.in}
\begin{abstract}
Despite the complexity of quantum systems in the real world,
models with just a few effective many-body states often suffice to describe their quantum dynamics, provided 
decoherence is accounted for.
We show that a machine learning algorithm is able to construct such models, given a straightforward set of quantum dynamics measurements. The effective Hilbert space can be a black box, with variations of the coupling to just one accessible output state being sufficient to generate the required training data.  We demonstrate through simulations of a Markovian open quantum system that a neural network can automatically detect the number $N $ of effective states and the most relevant Hamiltonian terms and state-dephasing processes and rates. For systems with $N\leq5$ we find typical mean relative errors of predictions in the $10 \%$ range.
With more advanced networks and larger training sets, it is conceivable that a future single software can provide the automated first stop solution to model building for an unknown device or system, complementing and validating the conventional approach based on physical insight into the system.
	\end{abstract}

	\maketitle
	\ssection{Introduction}
	\label{NN_NN_NN_intro}
%
	Simulating complex large quantum systems poses one of the most difficult outstanding computational problems. Fortunately, the dynamics of physical systems or devices can often be understood by reducing the very large many-body Hilbert space to just a few effective states that are crucial for a system. Knowing the Hamiltonian in this subspace can then enable practical computations, if
dephasing due to less important states of the device is incorporated \cite{Schlosshauer_decoherence_review}. 
For example light-harvesting complexes can be described considering a basis of excitations localized on individual pigment molecules, 
with decoherence arising from a bath of internal molecular vibrations \cite{engel2007evidence,calvin1983artificial,saikin:excitonreview,ReMaKue01_137_},
and photophysics in organic-semiconductors can be captured using just a few selected exciton and charge transfer states
\cite{Maimaris_excitons_organic_optoelec_NatComm}, with decoherence due to a phonon environment. Similar models can describe
bulk semi-conductors \cite{wheeler2013exciton} or superconducting circuits  \cite{mostame2012quantum}.
Constructing models to describe experiments conventionally requires the identification of a suitable effective space with calculation or measurements of Hamiltonian matrix elements (MEs) and dephasing rates.
	
Here, we present an approach for automatically creating quantum models leveraging machine learning (ML), which is emerging as a versatile tool for analyzing open system quantum dynamics, see e.g.~\cite{hase2017machine,papivc2022neural,luo2022autoregressive,bandyopadhyay2018applications,gentile2021learning}. Our algorithm can learn from a restrictive set of data, obtainable by simple measurements: (i) a random input state is prepared at time $t=0$, (ii) quantum dynamics involves a black-box region of Hilbert space, inaccessible to direct measurements, (iii) the population of just a single accessible output state is measured after a fixed delay time $t^*$.
The physical structure of the inaccessible black-box can be learned from those measurements, provided the coupling between black-box region and output state can be systematically varied to generate large datasets. Physical tuning knobs for this could include varying strain, temperature, resonance conditions or movement of subunits. Our algorithm can, to a certain extent, even overcome incomplete knowledge regarding the coupling to the output state and its variations. 

	\begin{figure}[htb!]
		\centering
		\includegraphics[width=1.05\linewidth]{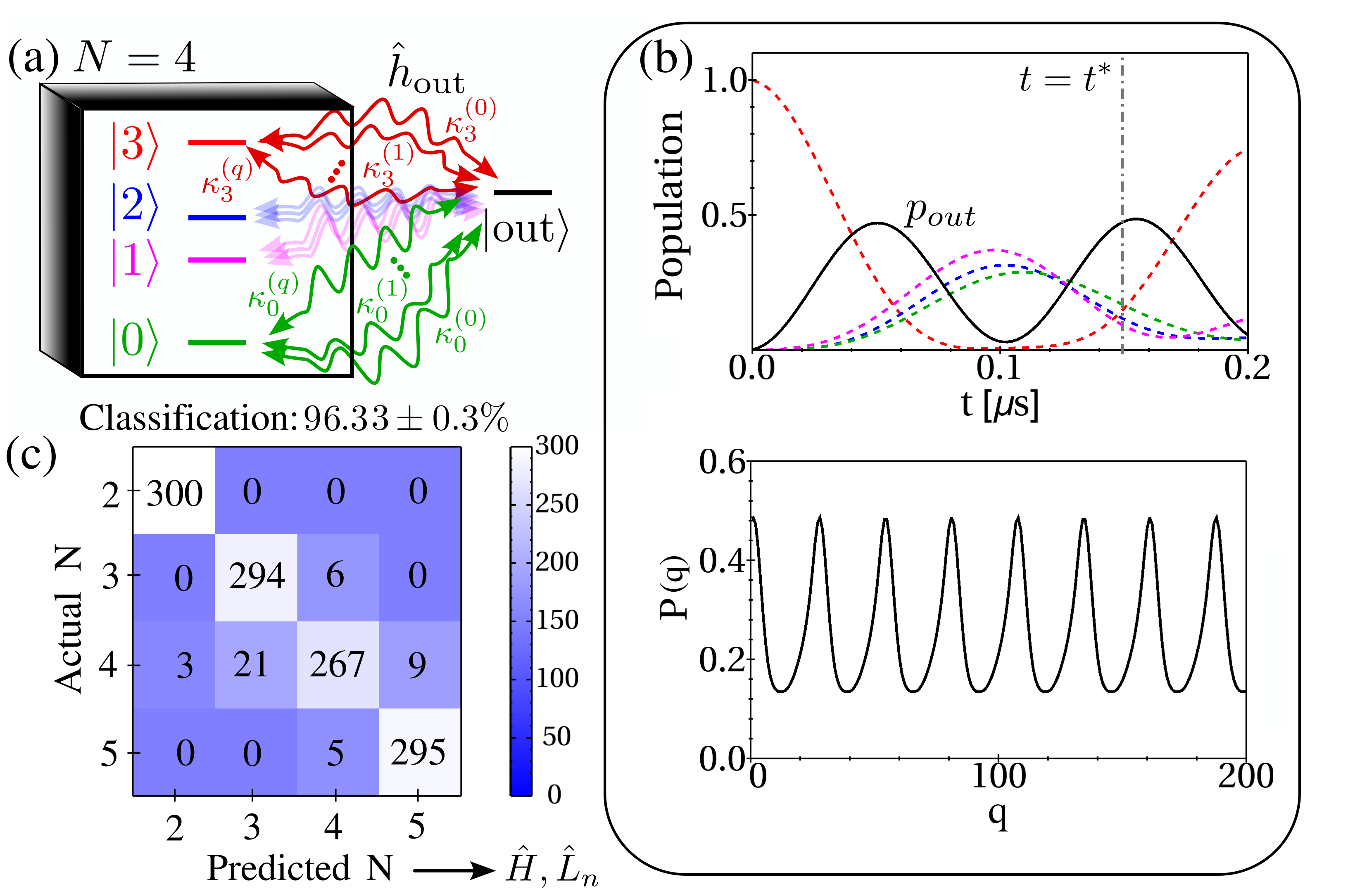}
		\caption{
Machine learning of effective state number. (a) Schematic of black box region in Hilbert space, here with $N=4$ effective states. A variable coupling to an accessible output state $\ket{\mbox{out}}$ with Hamiltonian $\sub{\hat{h}}{out}$ is used to generate large data sets. (b) (\textit{top}) Exemplary evolution of populations for the five states, $P_n(t)$.
(\textit{bottom}) We extract $\sub{P}{out}(t^*)$, at the time $t^*$ shown as dot-dashed vertical grey line in the top panel, for $200$ varied realisations $\sub{\hat{h}}{out}^{(q)}$ indexed by $q$. (c) The latter data is provided to a machine learning algorithm to extract the number of states $N$. The 2D histogram shows the number of testcases with given predicted and actual number of states $N$, hence the anti-diagonal counts correct classification. Based on $N$, we later demonstrate extraction of the Hamiltonian $\hat{H}$ and Lindblad operators $\hat{L}$.	\label{fig1_sketch}}
	\end{figure}
	As a proof-of-concept in the Markovian case, we train a neural network on numerical solutions of local Lindblad master equations for few-state systems, to recognise first the number of effective states, then based on that all MEs $H_{nm}$ of a real randomized Hamiltonian $\hat{H}$ and randomized Lindblad operators $\hat{L}_n$.
We reach mean relative errors of approximately $10\%$ for MEs $H_{nm}$ and $8\%$ for the dephasing rates in $\hat{L}_n$, using up to $N\leq 4$ effective states. Trained on simulations for a diverse range of parameters, the network may also be fed with experimental data and will supply viable effective state models with all ingredients for the unitary and non-unitary description of the experiment. These models can then guide more precise follow ups incorporating physical insight, could constrain parameter choices for those and can narrow down error sources in quantum device manufacturing.

A conceptually related approach is quantum process tomography (QPT) \cite{Poyatos_QPT,Chuang_Nielsen_QPT}, which reconstructs how a black-box maps an input to an output Hilbert space. QPT necessitates input and output state measurements in multiple different bases \cite{Torlai_QPT_ML} and does not provide all details of the interim black-box dynamics, but only the net mapping from input to output. In contrast, our ML architecture can provide unitary and non-unitary elements within the black-box, and only requires measurements of one population element, in one basis. Related proposals to leverage ML in the context of quantum dynamics include simulating open quantum systems \cite{bandyopadhyay2018applications,luo2022autoregressive}, quantum state tomography \cite{torlai2018neural,torlai2019integrating} and quantum network tomography \cite{de2022quantum}.
	
	\ssection{System and model}
	\label{sec:NN_system}
	%
	We consider a small quantum many-body system, the dynamics of which can be described
	by $N+1$ effective states $\{\ket{0},\cdots,\ket{N-1},\ket{\mbox{out}} \}$ within the many-body Hilbert space. Most of these states are considered part of a black-box, they are relevant during dynamics but are not accessible by measurements, except for the state $\ket{\mbox{out}}$, of which we can measure the population. States $\ket{n}$ in the black-box and transitions between them are affected by an environment, yielding an open quantum system. We envisage that both the system and its environment arise as parts of a mesoscopic quantum system, such as a nano-scale device.
	
	We assume that within the effective state space the system evolves according to the Lindblad equation ($\hbar=1$)
	\begin{equation}
		\dot{\hat{\rho}}=-i[\hat{H},\hat{\rho}]+\sum_{n=0}^{N-1}{\cal L}_{\hat{L}_n}[\hat{\rho}],
		\label{NN_Lindblad}
	\end{equation}
	for the density matrix $\hat{\rho}=\sum_{nm}\rho_{nm}\ket{n}\bra{m}$,
	with ${\cal L}_{\hat{L}}[\hat{O}]=\hat{L}\hat{O}\hat{L}^\dagger- \{\hat{L}^\dagger\hat{L}/2,\hat{O}\}$, and $\{,\}$ the anti-commutator.
	The Hamiltonian is 
	\begin{eqnarray}
		\hat{H}&=&\sum_{n,m=0}^{N-1}H_{nm}\ket{n}\bra{m}+\hat{h}^{(q)}_{out},
		\label{NN_Hprime}
	\end{eqnarray}
	with randomised matrix elements $H_{nm}$ in the black box Hilbert space excluding the output site, 
	and an output coupling $\hat{h}^{(q)}_{out}=\sum_{n=0}^{N-1} \kappa_{n}^{(q)} [\ket{n}\bra{\mbox{out}} +c.c.]$,
	which is deterministically tunable, parametrized by an integer tuning index $q$. See SI for more information.
	Within the black box, effective measurements of state populations by the environment cause~dephasing.
 This is captured by Lindblad operators $\hat{L}_n=\Gamma_n\ket{n}\bra{n}$ with random rates $\Gamma_n$. Both sets of MEs, $H_{nm}$ and ($\Gamma_n$), are drawn from uniform distributions between $0$ and $\sub{H}{max}$ ($\sub{\Gamma}{max}$). Dephasing operators do not involve $\ket{out}$.
 
 The objective of the machine learning algorithm is now to propose a valid small physical model, by infering $H_{nm}$ and $\Gamma_n$, based solely on measurements of the population in the output state, $\sub{P}{out}(t^*)=\mbox{Tr}[\hat{\rho}(t^*) \ket{\mbox{out}}\bra{\mbox{out}}]$, at a fixed time $t^*$. To generate the required large data-sets for training a neural network we will exploit the output coupling tuning indexed by $q$.
 The algorithm then operates in two stages: 
	
	\ssection{Effective Hilbert space extraction}
	\label{sec:Hilbert space}
	%
	Firstly, in the classification stage, the algorithm will infer the required number of effective states $N$. 
	The system is initialized in $\ket{1}$, and is then evolved using \eref{NN_Lindblad}. In \frefp{fig1_sketch}{b}, we show a typical time-evolution of the population in all states $P_n(t)=|\braket{n}{\Psi(t)}|^2$ (dashed lines), highlighting the measurable transition probability to the output state $\ket{\mbox{out}}$ (black-solid). At time $t^*=0.15$ $\mu$s, indicated with the vertical dot-dashed line in the top-panel, we then sample $P(q)=\sub{P}{out}(\kappa_{n}^{(q)},N,\hat{H},\hat{L};t_{end})$.  Through their dependence on $\kappa_{n}^{(q)}$ we generate one datapoint for each of the $Q=200$ different instances $q$. An exemplary variation of $P(q)$ is shown in the bottom panel of \frefp{fig1_sketch}{b}. Populations $P(q)$ are then used as input for the K-nearest neighbor (K-NN) scheme \cite{mucherino2009k}, to classify the system based on the required number of states $N$. For training purposes, we considered $D=10^4$ such datasets in total, with $2500$ datasets for each of $N\in\{2,3,4,5\}$, while $H_{nm}=H_{mn}$ and $\Gamma_n$ are uniformly distributed, random, positive real numbers below $\sub{H}{max}=\sub{\Gamma}{max}=1$ MHz. Here we arbitrarily choose $1$ MHz as characteristic energy scale of the problem. We then test the algorithm on $1200$ additional datasets, for which we show the performance of the K-NN classifier in \frefp{fig1_sketch}{c}. In the matrix histogram, anti-diagonal entries number the correct classifications, the figure thus implies successful classification for $2$ to $5$ dimensional effective state spaces, with an accuracy for finding the correct $N$ of $96.33\pm0.3\%$, averaged over all $N$.
	\begin{figure}[htb]
		\centering
		\includegraphics[width=1.\linewidth]{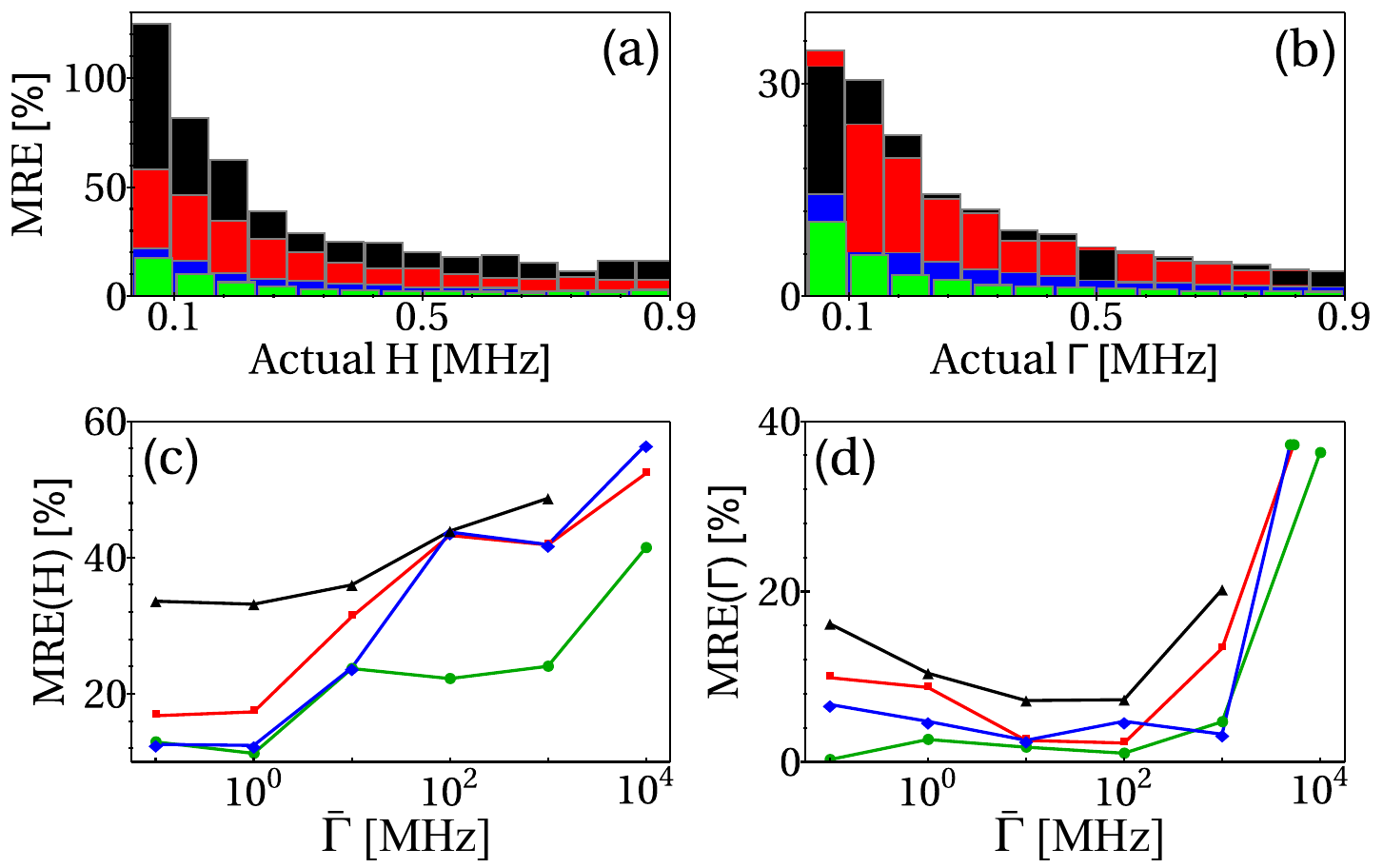}
		\caption{Performance of automated model building for different system sizes and values of matrix elements.
		(a) Histograms of the mean relative error (MRE) in $H_{nm}$ and (b) $\Gamma_n$ for $N=2$ (green), $N=3$ (blue), $N=4$ (red) and $N=5$ (black) as a function of the actual MEs or dephasing rates to be reconstructed. Average MRE for (c) $H_{nm}$ and (d) $\Gamma_n$ for varying mean dephasing rates $\bar{\Gamma}$, while the mean of all Hamiltonian MEs is kept constant at $\overline{H}=0.5$ MHz, and the training data size at $D=10^4$ elements.
		 \label{fig_model_generation}}
	\end{figure}
	%
	
	\ssection{Model generation}
	\label{sec:model_generation}
	%
Building on the classification result $N$, the algorithm subsequently infers the matrix elements of the effective Hamiltonian $H_{nm}$ and dephasing rates $\Gamma_n$, using an artificial neural network (ANN) for regression. Similar to the classification step, we provide output state populations $P(q)$ at a fixed time as training data to the ANN, while varying the couplings $\kappa_n^{(q)}$ with that output state. Details of the network architecture are provided in the SI.
	
\begin{figure}[tb]
		\centering
		\includegraphics[width=0.99\linewidth]{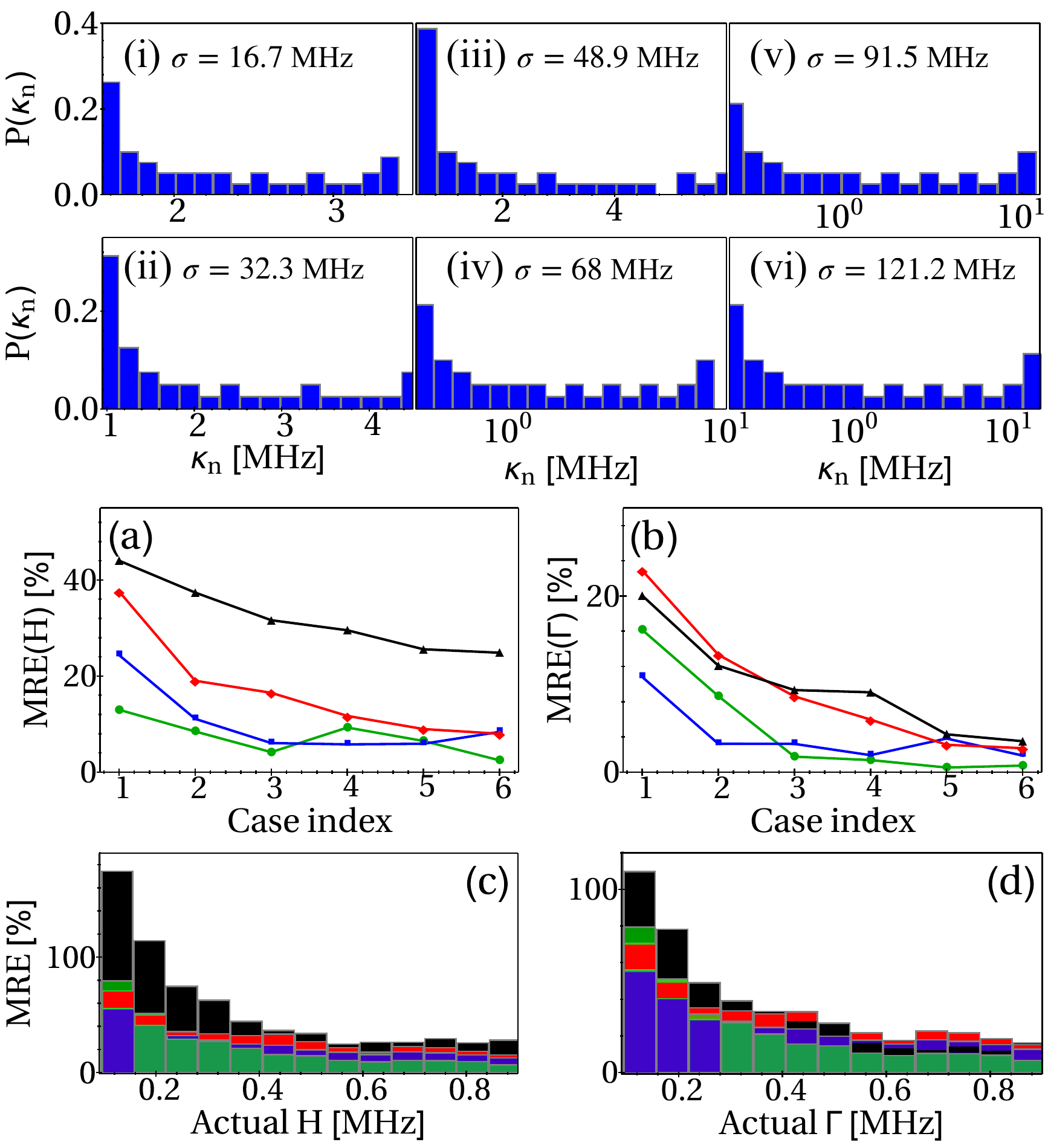}
		\caption{Results of the ML algorithm with unknown output couplings. (i-vi) Histogram of the distribution of the output couplings $\kappa_{n}^{(q)}$ in six different cases numbered by the panel index. Note the varying abscissae, with some being logarithmic. (a) MRE of $H_{nm}$ and (b) of $\Gamma_n$ for training and testing on the distributions of output couplings with case index (i-vi) for $N=2$ (green) $N=3$ (blue), $N=4$ (red) and $N=5$ (black). (c) The corresponding histograms of MRE in $H_{nm}$ and (d) $\Gamma_n$ for joint training on cases $\{$iv,v,vi$\}$, but testing on all six cases simultaneously. 
			\label{fig_alpha}}
	\end{figure}

	In \fref{fig_model_generation} we evaluate the performance of the algorithm under various conditions for $N\leq5$, using
	the mean relative error (MRE) as a performance metric. For the MRE we first average the relative deviation between predicted and actual ME over the Hamiltonian as $\mathrm{MRE}_q(H)=\frac{1}{N^2}\sum_{nm}(|H_{nm}^{\mathrm{predicted}}-H_{nm}^{\mathrm{actual}}|)/H_{nm}^{\mathrm{actual}}$, for each realisation $q$, and then average these over all realisations, denoting the average by $\mathrm{MRE}=\overline{\mathrm{MRE}_q}$ (similarly for $\hat{L}$).
	For $N=(2,3,4)$ we find a MRE of $(4.2\%,6\%,16.5\%)$ for the random Hamiltonian ($H_{nm}$) and $(2\%,3.6\%,9.6\%)$ for dephasing rates ($\Gamma_{n}$), with typical relative deviation of $10\%$ and $5\%$, respectively. The Hamiltonian and dephasing rates are inferred simultaneously, based on the distribution of output coupling shown in \fref{fig_alpha}~(iii). 
We can see in \frefp{fig_model_generation}{a,b} that the MRE is smaller for larger MEs, owing to the mean absolute error MAE$=\overline{|\text{predicted}-\text{actual}|}$ being nearly independent of energy, with MAE=$(1.7,2.2,6.5)\times 10^{-2}$ MHz for Hamiltonian MEs and MAE=$(0.6,1.3,3.6)\times 10^{-2}$ MHz for dephasing rates. The MAE in turn increases with $N$, $\overline{H}\equiv \sum_{n,m} H_{nm}/N^2$ and $\overline{\Gamma}\equiv \sum_{n} \Gamma_n/N$.
The Algorithm is clearly challenged by larger Hilbertspace sizes, with the MRE reaching $32\%$ for $N=5$ if we restrict ourselves to training data-sets with $D=10^4$ elements.
However, we find that this can again be reduced to MRE=$4.9\%$ for the $H_{nm}$ (MRE=$17\%$ for $\Gamma$) by increasing the 
 training data size ($D=10^5$), which in turn challenges our computational resources.
	
To further assess how the performance of model generation depends on the overall dephasing strength, we vary the mean dephasing rate $\overline{\Gamma}$ from $10^{-1}$ to $10^4$ MHz, while keeping $H_{nm}\in [0,\:1]$ MHz.
We find that performance of predicting $H_{nm}$ degrades as dephasing becomes stronger, shown in \frefp{fig_model_generation}{c}.
While the algorithm infers $H_{nm}$ well up to $\overline{\Gamma} \approx \overline{H} \sim {\cal O}(1$ MHz$)$, the MRE increases rapidly for $\overline{\Gamma}\gg \overline{H}$, due to the fast relaxation to a steady state, which reduces information contained in interference features. 
In contrast, the performance of predicting $\Gamma_{n}$ itself (\frefp{fig_model_generation}{d}) shows a more complex non-monotonic dependence on $\overline{\Gamma}$ and remains acceptable up to larger dephasing strengths $\overline{\Gamma} \approx 10^3$, since the steady states themselves strongly depend on $\overline{\Gamma}$. For even stronger dephasing, all configurations $q$ result in the same steady-state and predictions fail, as discussed in more detail in \cite{mukherjeeNN:long}.

In summary, the algorithm works well for $N<5$ with $D=10^4$ training samples, with indications that larger $N$ simply require larger $D$. Further improvements may be possible by leveraging more complex neural networks or quantum neural networks with quantum neurons \cite{schuld2014quest}.
	
	\ssection{Model generation for unknown output coupling}
	\label{sec:alpha_output}
	%
The objective of our ML algorithm is to infer effective few-state models for unknown or incompletely characterised  systems from restricted measurements of  dynamics traversing the system. We have so far demonstrated that this is possible, albeit using specific sets of varied output couplings $\kappa_{n}^{(q)}$ which were \emph{identical} between training and test sets. For the practical use of a pre-trained network on an unknown quantum system, this would necessitate a-priori knowledge of how the coupling to output states changes when a certain experimental parameter is varied. While the capability to infer some part of the Hamiltonian subject to the condition that another part is already known may also be useful, we now provide evidence that this limitation can be overcome. 

We find that model generation can be successful when the network has been trained on many different sets of varied output coupling, even if the one actually used in the test data has not been included in these sets. This is demonstrated in \fref{fig_alpha}. We consider six random distributions for $\kappa_{n}^{(q)}$ with very different standard deviations
$\sigma=\{16.7,32.3,48.9,68,91.5,121.2\}$ MHz and skewness $\mu_3=\{2.1,2.8,3.5,4.2,4.7,5.2\}\times 10^{-2}$ MHz$^2$, as shown in \fref{fig_alpha}~(i-vi). 
The origin of these is discussed in the SI.
We first train and test the network using each output coupling distribution separately, to explore how their statistical properties affect performance, and present the results in \fref{fig_alpha}~(a-b). The MRE decreases for cases with a larger standard deviation, presumably since these provide more distinctive information to the network.
Going one step further, inspired by \rref{bhavna:NNridges}, we trained \emph{a single} neural network on a mix of cases iv-vi and then selected test data from the complete mixture of distributions (cases i-vi). The test data thus included output coupling data sets that the algorithm did not encounter before. The results are shown in \fref{fig_alpha}~(c-d), for $N=2,3$ and $4$. They indicate decent performance even in this challenging case, with a characteristic MRE of about $12\%$ for most MEs. Even if the test-data is fully restricted to the unseen sets i-iii, the error increases only by a factor of two, keeping at least $N=2,3$ viable. This suggests that a neural network which has been trained on a wide range of characteristically different output coupling sets can build a model without  prior information of the true output couplings for an unknown device or system.

	%
	\ssection{Outlook and conclusions}
	\label{sec:outlook}
	%
We have demonstrated that a machine learning architecture can infer viable quantum dynamical models consisting of a few relevant effective states based on restricted information such as the population in a single selected ``output'' state after a fixed time of evolution. This process does not require any prior knowledge of the system, only access to training data generated by varying some system parameter(s) in simulations or experiments. Here we considered small systems with $2\leq N\leq5$ effective states, for which we could show typical relative errors of model parameters in the 10\% range. It is conceivable that with much larger computational resources for network training, larger systems will also be tractable or lower errors reachable. 
	
While model building by machine learning may also be useful to analyse data from complex simulations, for example in material science or quantum chemistry, its most powerful application would be to interpret experiments. To this end, one can train the network on a very large set of simulation data, including even more diverse variations of the unknown output coupling $\kappa_{n}^{(q)}$ than we explored here. It can then generate effective models based on measurements of quantum dynamics ($P(q)$), for example in supra-molecular structures \cite{peng2015biological}, molecular wires \cite{schwartz2003conjugated} or organic semi-conductors \cite{Maimaris_excitons_organic_optoelec_NatComm}.
	
Machine learning model building can complement the arsenal of traditional model building based on physical insights by providing an initial starting point of parameter determination, validating manually constructed models, cross-checking parameters or perhaps providing guidance regarding initially undiscovered relevant states.
	
Our demonstration considered a quite general case with all MEs of the system Hamiltonian and all dephasing strengths unknown. Any prior knowledge of the quantum system constraining some of these MEs ought to significantly enhance the performance of the algorithm. While we only considered simple local dephasing and a Markovian system-environment model, the approach might be extendable towards more comple dephasing processes, or inferring also relaxation rates or non-Markovian environmental properties, i.e.~spectral densities.
A key advantage of the present approach over more complex designs with similar objectives \cite{Heightman_hamillearn_neuraldiffeq,carleo2017solving,bertalan2019learning} is the simplicity of the ML formalism and input required.

\acknowledgments
We acknowledge interesting discussions with Abhijit Pendse, Ritesh Pant and thank the Max-Planck society under the MPG-IISER partner group program as well as the Indo-French Centre for the Promotion of Advanced Research - CEFIPRA for financial support. K.M.~acknowledges the Indian Ministry of Education for the Prime Minister's Research Fellowship (PMRF).
We acknowledge support by the European Union's Horizon Europe research and innovation program under the project MLQ under Marie Skłodowska-Curie grant agreement number 101120240. Work was also supported by the CNRS through the EMERGENCE@INC2024 project DINOPARC and by the French National Research Agency under the Investments of the Future Program project ANR-21-ESRE-0032 (aQCess) and the Institut Universitare de France
(IUF).

	\appendix
	\beginsupplement
\section{Supplemental material}
	\begin{figure}[htb]
		\centering
		\includegraphics[width=0.99\columnwidth]{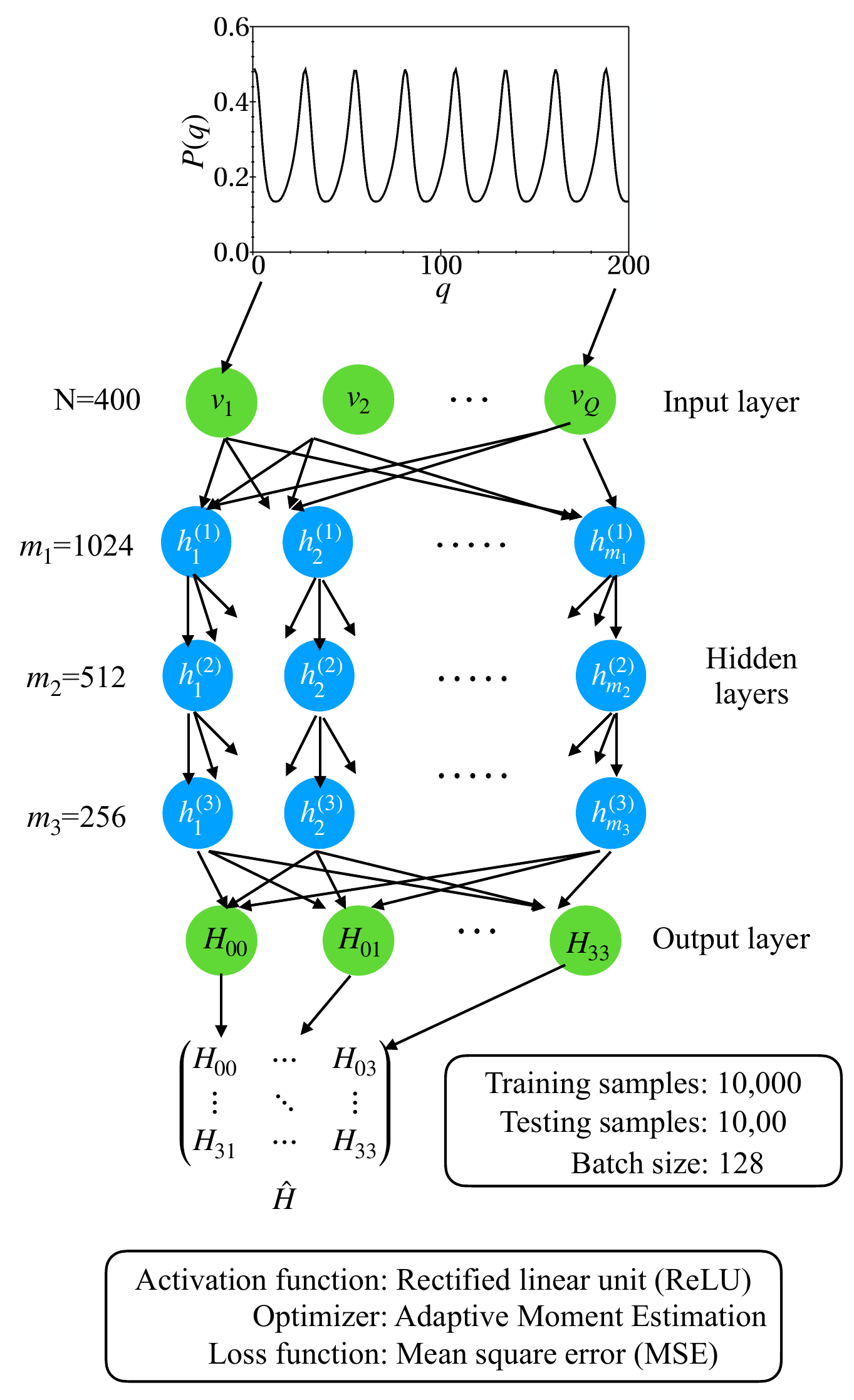}
		\caption{Schematic representation of artificial neural network (ANN) with an
			input layer consisting of $N=400$ input neurons ($\mathbf{v}$), $L=3$ hidden layers consisting of $m_l$ hidden
			neurons ($\mathbf{h}^{(l)}$) each and an output layer with $K$ output neurons (e.g.~$H_{nm}$). Information is processed
			from top ($P(q)$) to bottom, as indicated by the arrows.}
		\label{NN_structure}
	\end{figure}
	%
	\section{Design of the Artificial Neural Network}
	\label{app:NN_design}
	%
	An artificial neural network with multi-target regression has been used to predict the matrix elements of the arbitrary Hamiltonian and decay operators. In general, an artificial neural network consists of three sections, for which we are using the following specifications:
	\begin{enumerate}
		\item Input layer: Our input layer consists of $Q=200$ neurons indexed by $q$, where the measurement data of ouput state populations, $P(q)=\sub{p}{out}(\kappa^{(q)}_{n},N,\hat{H},\hat{L};t^*)$ is provided.
		\item Hidden layer: We have typically included $L=3$ hidden layers, where the $i$th layer in the forward direction has $2^{11-i}$ neurons. All the neurons are connected to each other with dropout = 0, and activated using a Rectified linear unit (ReLU) activation function. We have used the mean squared error (MSE) as our loss function and Adaptive Moment estimation (ADAM) as optimizer.
		\item Output layer: We vary the number of neurons at the output layer based on data type required for the prediction. For instance, to predict the MEs of Hamiltonian of $(3\times 3)$, we need $9$ real numbers, requiring $9$ neurons at the output layer.
	\end{enumerate}

	The neural network configured like this was trained with $10^4$ training datasets in all cases for typically $10^3$ epochs (iterations for the loss function optimisation during training).
	To prevent underfitting or overfitting, the number of epochs was adjusted from that value case by case. 
	
	\section{Variation of output coupling}
	\label{app:details_on_powerlaws}
	
	The output couplings $\kappa_{n}^{(q)}$ in \eref{NN_Hprime} and shown in \fref{fig_alpha}(1-6) of the main text 
	have been realized using fictitious spin-exchange interactions in the form
	\begin{equation}
		\kappa^{(q)}_{n} =\frac{J_{n,out}}{|r_{n,q}|^\alpha},
	\end{equation} 
	where $J_{n,out}$ are interaction strengths, $r_{n,q}=|\mathbf{x}_n - \sub{\mathbf{x}}{out}(q)|$ is the distance between a fictitious positon $\mathbf{x}_n$ allocated to state $\ket{n}$ and $\sub{\mathbf{x}}{out}(q)$ allocated to the output state, while $\alpha\in\{1,2,\cdots,6\}$ sets the power-law of this fictitious interaction and $q$ parametrizes the changeable output state ``location''. 
We use power laws with $\alpha$ equal to the case index in \fref{fig_alpha} to be specific.

We refer to all the above as ``fictitious'' since the downstream approach makes no reference to any positions $\mathbf{x}$ and the states $\ket{n}$ are all many-body states that do not necessarily carry any spatial degrees of freedom. Instead we envisage that tuning of $\kappa^{(q)}_{n}$ with some control parameter $q$ can be generated 
by adjusting e.g.~strain, temperature or resonance conditions in some device or material. For the processing with our NN algorithm, all that would matter is that a set of different interaction strengths $\{ \kappa^{(q)}_{n}\}$ between all states $\ket{n}$ and $\ket{out}$ is generated.


\begin{thebibliography}{28}%
	\makeatletter
	\providecommand \@ifxundefined [1]{%
		\@ifx{#1\undefined}
	}%
	\providecommand \@ifnum [1]{%
		\ifnum #1\expandafter \@firstoftwo
		\else \expandafter \@secondoftwo
		\fi
	}%
	\providecommand \@ifx [1]{%
		\ifx #1\expandafter \@firstoftwo
		\else \expandafter \@secondoftwo
		\fi
	}%
	\providecommand \natexlab [1]{#1}%
	\providecommand \enquote  [1]{``#1''}%
	\providecommand \bibnamefont  [1]{#1}%
	\providecommand \bibfnamefont [1]{#1}%
	\providecommand \citenamefont [1]{#1}%
	\providecommand \href@noop [0]{\@secondoftwo}%
	\providecommand \href [0]{\begingroup \@sanitize@url \@href}%
	\providecommand \@href[1]{\@@startlink{#1}\@@href}%
	\providecommand \@@href[1]{\endgroup#1\@@endlink}%
	\providecommand \@sanitize@url [0]{\catcode `\\12\catcode `\$12\catcode
		`\&12\catcode `\#12\catcode `\^12\catcode `\_12\catcode `\%12\relax}%
	\providecommand \@@startlink[1]{}%
	\providecommand \@@endlink[0]{}%
	\providecommand \url  [0]{\begingroup\@sanitize@url \@url }%
	\providecommand \@url [1]{\endgroup\@href {#1}{\urlprefix }}%
	\providecommand \urlprefix  [0]{URL }%
	\providecommand \Eprint [0]{\href }%
	\providecommand \doibase [0]{https://doi.org/}%
	\providecommand \selectlanguage [0]{\@gobble}%
	\providecommand \bibinfo  [0]{\@secondoftwo}%
	\providecommand \bibfield  [0]{\@secondoftwo}%
	\providecommand \translation [1]{[#1]}%
	\providecommand \BibitemOpen [0]{}%
	\providecommand \bibitemStop [0]{}%
	\providecommand \bibitemNoStop [0]{.\EOS\space}%
	\providecommand \EOS [0]{\spacefactor3000\relax}%
	\providecommand \BibitemShut  [1]{\csname bibitem#1\endcsname}%
	\let\auto@bib@innerbib\@empty
	\bibitem [{\citenamefont
		{Schlosshauer}(2005)}]{Schlosshauer_decoherence_review}%
	\BibitemOpen
	\bibfield  {author} {\bibinfo {author} {\bibfnamefont {M.}~\bibnamefont
			{Schlosshauer}},\ }\bibfield  {title} {\bibinfo {title} {Decoherence, the
			measurement problem, and interpretations of quantum mechanics},\ }\href@noop
	{} {\bibfield  {journal} {\bibinfo  {journal} {Rev. Mod. Phys.}\ }\textbf
		{\bibinfo {volume} {76}},\ \bibinfo {pages} {1267} (\bibinfo {year}
		{2005})}\BibitemShut {NoStop}%
	\bibitem [{\citenamefont {Engel}\ \emph {et~al.}(2007)\citenamefont {Engel},
		\citenamefont {Calhoun}, \citenamefont {Read}, \citenamefont {Ahn},
		\citenamefont {Man{\v{c}}al}, \citenamefont {Cheng}, \citenamefont
		{Blankenship},\ and\ \citenamefont {Fleming}}]{engel2007evidence}%
	\BibitemOpen
	\bibfield  {author} {\bibinfo {author} {\bibfnamefont {G.~S.}\ \bibnamefont
			{Engel}}, \bibinfo {author} {\bibfnamefont {T.~R.}\ \bibnamefont {Calhoun}},
		\bibinfo {author} {\bibfnamefont {E.~L.}\ \bibnamefont {Read}}, \bibinfo
		{author} {\bibfnamefont {T.-K.}\ \bibnamefont {Ahn}}, \bibinfo {author}
		{\bibfnamefont {T.}~\bibnamefont {Man{\v{c}}al}}, \bibinfo {author}
		{\bibfnamefont {Y.-C.}\ \bibnamefont {Cheng}}, \bibinfo {author}
		{\bibfnamefont {R.~E.}\ \bibnamefont {Blankenship}},\ and\ \bibinfo {author}
		{\bibfnamefont {G.~R.}\ \bibnamefont {Fleming}},\ }\bibfield  {title}
	{\bibinfo {title} {Evidence for wavelike energy transfer through quantum
			coherence in photosynthetic systems},\ }\href@noop {} {\bibfield  {journal}
		{\bibinfo  {journal} {Nature}\ }\textbf {\bibinfo {volume} {446}},\ \bibinfo
		{pages} {782} (\bibinfo {year} {2007})}\BibitemShut {NoStop}%
	\bibitem [{\citenamefont {Calvin}(1983)}]{calvin1983artificial}%
	\BibitemOpen
	\bibfield  {author} {\bibinfo {author} {\bibfnamefont {M.}~\bibnamefont
			{Calvin}},\ }\bibfield  {title} {\bibinfo {title} {Artificial photosynthesis:
			Quantum capture and energy storage},\ }\href@noop {} {\bibfield  {journal}
		{\bibinfo  {journal} {Photochemistry and Photobiology}\ }\textbf {\bibinfo
			{volume} {37}},\ \bibinfo {pages} {349} (\bibinfo {year} {1983})}\BibitemShut
	{NoStop}%
	\bibitem [{\citenamefont {Saikin}\ \emph {et~al.}(2013)\citenamefont {Saikin},
		\citenamefont {Eisfeld}, \citenamefont {Valleau},\ and\ \citenamefont
		{{Aspuru-Guzik}}}]{saikin:excitonreview}%
	\BibitemOpen
	\bibfield  {author} {\bibinfo {author} {\bibfnamefont {S.~K.}\ \bibnamefont
			{Saikin}}, \bibinfo {author} {\bibfnamefont {A.}~\bibnamefont {Eisfeld}},
		\bibinfo {author} {\bibfnamefont {S.}~\bibnamefont {Valleau}},\ and\ \bibinfo
		{author} {\bibfnamefont {A.}~\bibnamefont {{Aspuru-Guzik}}},\ }\bibfield
	{title} {\bibinfo {title} {Photonics meets excitonics: natural and artificial
			molecular aggregates},\ }\href@noop {} {\bibfield  {journal} {\bibinfo
			{journal} {Nanophotonics}\ }\textbf {\bibinfo {volume} {2}},\ \bibinfo
		{pages} {21} (\bibinfo {year} {2013})}\BibitemShut {NoStop}%
	\bibitem [{\citenamefont {Renger}\ \emph {et~al.}(2001)\citenamefont {Renger},
		\citenamefont {May},\ and\ \citenamefont {K\"uhn}}]{ReMaKue01_137_}%
	\BibitemOpen
	\bibfield  {author} {\bibinfo {author} {\bibfnamefont {T.}~\bibnamefont
			{Renger}}, \bibinfo {author} {\bibfnamefont {V.}~\bibnamefont {May}},\ and\
		\bibinfo {author} {\bibfnamefont {O.}~\bibnamefont {K\"uhn}},\ }\bibfield
	{title} {\bibinfo {title} {Ultrafast excitation energy transfer dynamics in
			photosynthetic pigment-protein complexes},\ }\href@noop {} {\bibfield
		{journal} {\bibinfo  {journal} {Physics Reports}\ }\textbf {\bibinfo {volume}
			{343}},\ \bibinfo {pages} {137} (\bibinfo {year} {2001})}\BibitemShut
	{NoStop}%
	\bibitem [{\citenamefont {Maimaris}\ \emph {et~al.}(2022)\citenamefont
		{Maimaris}, \citenamefont {Pettipher}, \citenamefont {Azzouzi}, \citenamefont
		{Walke}, \citenamefont {Zheng}, \citenamefont {Gorodetsky}, \citenamefont
		{Dong}, \citenamefont {Tuladhar}, \citenamefont {Crespo}, \citenamefont
		{Nelson}, \citenamefont {Tisch},\ and\ \citenamefont
		{Bakulin}}]{Maimaris_excitons_organic_optoelec_NatComm}%
	\BibitemOpen
	\bibfield  {author} {\bibinfo {author} {\bibfnamefont {M.}~\bibnamefont
			{Maimaris}}, \bibinfo {author} {\bibfnamefont {A.~J.}\ \bibnamefont
			{Pettipher}}, \bibinfo {author} {\bibfnamefont {M.}~\bibnamefont {Azzouzi}},
		\bibinfo {author} {\bibfnamefont {D.~J.}\ \bibnamefont {Walke}}, \bibinfo
		{author} {\bibfnamefont {X.}~\bibnamefont {Zheng}}, \bibinfo {author}
		{\bibfnamefont {A.}~\bibnamefont {Gorodetsky}}, \bibinfo {author}
		{\bibfnamefont {Y.}~\bibnamefont {Dong}}, \bibinfo {author} {\bibfnamefont
			{P.~S.}\ \bibnamefont {Tuladhar}}, \bibinfo {author} {\bibfnamefont
			{H.}~\bibnamefont {Crespo}}, \bibinfo {author} {\bibfnamefont
			{J.}~\bibnamefont {Nelson}}, \bibinfo {author} {\bibfnamefont {J.~W.~G.}\
			\bibnamefont {Tisch}},\ and\ \bibinfo {author} {\bibfnamefont {A.~A.}\
			\bibnamefont {Bakulin}},\ }\bibfield  {title} {\bibinfo {title} {Sub-10-fs
			observation of bound exciton formation in organic optoelectronic devices},\
	}\href@noop {} {\bibfield  {journal} {\bibinfo  {journal} {Nat. Commun.}\
		}\textbf {\bibinfo {volume} {13}},\ \bibinfo {pages} {4949} (\bibinfo {year}
		{2022})}\BibitemShut {NoStop}%
	\bibitem [{\citenamefont {Wheeler}\ and\ \citenamefont
		{Zhang}(2013)}]{wheeler2013exciton}%
	\BibitemOpen
	\bibfield  {author} {\bibinfo {author} {\bibfnamefont {D.~A.}\ \bibnamefont
			{Wheeler}}\ and\ \bibinfo {author} {\bibfnamefont {J.~Z.}\ \bibnamefont
			{Zhang}},\ }\bibfield  {title} {\bibinfo {title} {Exciton dynamics in
			semiconductor nanocrystals},\ }\href@noop {} {\bibfield  {journal} {\bibinfo
			{journal} {Adv. Mater.}\ }\textbf {\bibinfo {volume} {25}},\ \bibinfo {pages}
		{2878} (\bibinfo {year} {2013})}\BibitemShut {NoStop}%
	\bibitem [{\citenamefont {Mostame}\ \emph {et~al.}(2012)\citenamefont
		{Mostame}, \citenamefont {Rebentrost}, \citenamefont {Eisfeld}, \citenamefont
		{Kerman}, \citenamefont {Tsomokos},\ and\ \citenamefont
		{Aspuru-Guzik}}]{mostame2012quantum}%
	\BibitemOpen
	\bibfield  {author} {\bibinfo {author} {\bibfnamefont {S.}~\bibnamefont
			{Mostame}}, \bibinfo {author} {\bibfnamefont {P.}~\bibnamefont {Rebentrost}},
		\bibinfo {author} {\bibfnamefont {A.}~\bibnamefont {Eisfeld}}, \bibinfo
		{author} {\bibfnamefont {A.~J.}\ \bibnamefont {Kerman}}, \bibinfo {author}
		{\bibfnamefont {D.~I.}\ \bibnamefont {Tsomokos}},\ and\ \bibinfo {author}
		{\bibfnamefont {A.}~\bibnamefont {Aspuru-Guzik}},\ }\bibfield  {title}
	{\bibinfo {title} {Quantum simulator of an open quantum system using
			superconducting qubits: exciton transport in photosynthetic complexes},\
	}\href@noop {} {\bibfield  {journal} {\bibinfo  {journal} {New J. Phys.}\
		}\textbf {\bibinfo {volume} {14}},\ \bibinfo {pages} {105013} (\bibinfo
		{year} {2012})}\BibitemShut {NoStop}%
	\bibitem [{\citenamefont {H{\"a}se}\ \emph {et~al.}(2017)\citenamefont
		{H{\"a}se}, \citenamefont {Kreisbeck},\ and\ \citenamefont
		{Aspuru-Guzik}}]{hase2017machine}%
	\BibitemOpen
	\bibfield  {author} {\bibinfo {author} {\bibfnamefont {F.}~\bibnamefont
			{H{\"a}se}}, \bibinfo {author} {\bibfnamefont {C.}~\bibnamefont
			{Kreisbeck}},\ and\ \bibinfo {author} {\bibfnamefont {A.}~\bibnamefont
			{Aspuru-Guzik}},\ }\bibfield  {title} {\bibinfo {title} {Machine learning for
			quantum dynamics: deep learning of excitation energy transfer properties},\
	}\href@noop {} {\bibfield  {journal} {\bibinfo  {journal} {Chem. Sci.}\
		}\textbf {\bibinfo {volume} {8}},\ \bibinfo {pages} {8419} (\bibinfo {year}
		{2017})}\BibitemShut {NoStop}%
	\bibitem [{\citenamefont {Papi{\v{c}}}\ and\ \citenamefont
		{de~Vega}(2022)}]{papivc2022neural}%
	\BibitemOpen
	\bibfield  {author} {\bibinfo {author} {\bibfnamefont {M.}~\bibnamefont
			{Papi{\v{c}}}}\ and\ \bibinfo {author} {\bibfnamefont {I.}~\bibnamefont
			{de~Vega}},\ }\bibfield  {title} {\bibinfo {title} {Neural-network-based
			qubit-environment characterization},\ }\href@noop {} {\bibfield  {journal}
		{\bibinfo  {journal} {Phys. Rev. A}\ }\textbf {\bibinfo {volume} {105}},\
		\bibinfo {pages} {022605} (\bibinfo {year} {2022})}\BibitemShut {NoStop}%
	\bibitem [{\citenamefont {Luo}\ \emph {et~al.}(2022)\citenamefont {Luo},
		\citenamefont {Chen}, \citenamefont {Carrasquilla},\ and\ \citenamefont
		{Clark}}]{luo2022autoregressive}%
	\BibitemOpen
	\bibfield  {author} {\bibinfo {author} {\bibfnamefont {D.}~\bibnamefont
			{Luo}}, \bibinfo {author} {\bibfnamefont {Z.}~\bibnamefont {Chen}}, \bibinfo
		{author} {\bibfnamefont {J.}~\bibnamefont {Carrasquilla}},\ and\ \bibinfo
		{author} {\bibfnamefont {B.~K.}\ \bibnamefont {Clark}},\ }\bibfield  {title}
	{\bibinfo {title} {Autoregressive neural network for simulating open quantum
			systems via a probabilistic formulation},\ }\href@noop {} {\bibfield
		{journal} {\bibinfo  {journal} {Phys. Rev. Lett.}\ }\textbf {\bibinfo
			{volume} {128}},\ \bibinfo {pages} {090501} (\bibinfo {year}
		{2022})}\BibitemShut {NoStop}%
	\bibitem [{\citenamefont {Bandyopadhyay}\ \emph {et~al.}(2018)\citenamefont
		{Bandyopadhyay}, \citenamefont {Huang}, \citenamefont {Sun},\ and\
		\citenamefont {Zhao}}]{bandyopadhyay2018applications}%
	\BibitemOpen
	\bibfield  {author} {\bibinfo {author} {\bibfnamefont {S.}~\bibnamefont
			{Bandyopadhyay}}, \bibinfo {author} {\bibfnamefont {Z.}~\bibnamefont
			{Huang}}, \bibinfo {author} {\bibfnamefont {K.}~\bibnamefont {Sun}},\ and\
		\bibinfo {author} {\bibfnamefont {Y.}~\bibnamefont {Zhao}},\ }\bibfield
	{title} {\bibinfo {title} {Applications of neural networks to the simulation
			of dynamics of open quantum systems},\ }\href@noop {} {\bibfield  {journal}
		{\bibinfo  {journal} {Chem. Phys.}\ }\textbf {\bibinfo {volume} {515}},\
		\bibinfo {pages} {272} (\bibinfo {year} {2018})}\BibitemShut {NoStop}%
	\bibitem [{\citenamefont {Gentile}\ \emph {et~al.}(2021)\citenamefont
		{Gentile}, \citenamefont {Flynn}, \citenamefont {Knauer}, \citenamefont
		{Wiebe}, \citenamefont {Paesani}, \citenamefont {Granade}, \citenamefont
		{Rarity}, \citenamefont {Santagati},\ and\ \citenamefont
		{Laing}}]{gentile2021learning}%
	\BibitemOpen
	\bibfield  {author} {\bibinfo {author} {\bibfnamefont {A.~A.}\ \bibnamefont
			{Gentile}}, \bibinfo {author} {\bibfnamefont {B.}~\bibnamefont {Flynn}},
		\bibinfo {author} {\bibfnamefont {S.}~\bibnamefont {Knauer}}, \bibinfo
		{author} {\bibfnamefont {N.}~\bibnamefont {Wiebe}}, \bibinfo {author}
		{\bibfnamefont {S.}~\bibnamefont {Paesani}}, \bibinfo {author} {\bibfnamefont
			{C.~E.}\ \bibnamefont {Granade}}, \bibinfo {author} {\bibfnamefont {J.~G.}\
			\bibnamefont {Rarity}}, \bibinfo {author} {\bibfnamefont {R.}~\bibnamefont
			{Santagati}},\ and\ \bibinfo {author} {\bibfnamefont {A.}~\bibnamefont
			{Laing}},\ }\bibfield  {title} {\bibinfo {title} {Learning models of quantum
			systems from experiments},\ }\href@noop {} {\bibfield  {journal} {\bibinfo
			{journal} {Nature Physics}\ }\textbf {\bibinfo {volume} {17}},\ \bibinfo
		{pages} {837} (\bibinfo {year} {2021})}\BibitemShut {NoStop}%
	\bibitem [{\citenamefont {Poyatos}\ \emph {et~al.}(1997)\citenamefont
		{Poyatos}, \citenamefont {Cirac},\ and\ \citenamefont
		{Zoller}}]{Poyatos_QPT}%
	\BibitemOpen
	\bibfield  {author} {\bibinfo {author} {\bibfnamefont {J.~F.}\ \bibnamefont
			{Poyatos}}, \bibinfo {author} {\bibfnamefont {J.~I.}\ \bibnamefont {Cirac}},\
		and\ \bibinfo {author} {\bibfnamefont {P.}~\bibnamefont {Zoller}},\
	}\bibfield  {title} {\bibinfo {title} {Complete characterization of a quantum
			process: The two-bit quantum gate},\ }\href@noop {} {\bibfield  {journal}
		{\bibinfo  {journal} {Phys. Rev. Lett.}\ }\textbf {\bibinfo {volume} {78}},\
		\bibinfo {pages} {390} (\bibinfo {year} {1997})}\BibitemShut {NoStop}%
	\bibitem [{\citenamefont {Chuang}\ and\ \citenamefont
		{Nielsen}(1997)}]{Chuang_Nielsen_QPT}%
	\BibitemOpen
	\bibfield  {author} {\bibinfo {author} {\bibfnamefont {I.~L.}\ \bibnamefont
			{Chuang}}\ and\ \bibinfo {author} {\bibfnamefont {M.~A.}\ \bibnamefont
			{Nielsen}},\ }\bibfield  {title} {\bibinfo {title} {Prescription for
			experimental determination of the dynamics of a quantum black box},\
	}\href@noop {} {\bibfield  {journal} {\bibinfo  {journal} {J. Mod. Opt.}\
		}\textbf {\bibinfo {volume} {44}},\ \bibinfo {pages} {2455} (\bibinfo {year}
		{1997})}\BibitemShut {NoStop}%
	\bibitem [{\citenamefont {Torlai}\ \emph {et~al.}(2023)\citenamefont {Torlai},
		\citenamefont {Wood}, \citenamefont {Acharya}, \citenamefont {Carleo},
		\citenamefont {Carrasquilla},\ and\ \citenamefont {Aolita}}]{Torlai_QPT_ML}%
	\BibitemOpen
	\bibfield  {author} {\bibinfo {author} {\bibfnamefont {G.}~\bibnamefont
			{Torlai}}, \bibinfo {author} {\bibfnamefont {C.~J.}\ \bibnamefont {Wood}},
		\bibinfo {author} {\bibfnamefont {A.}~\bibnamefont {Acharya}}, \bibinfo
		{author} {\bibfnamefont {G.}~\bibnamefont {Carleo}}, \bibinfo {author}
		{\bibfnamefont {J.}~\bibnamefont {Carrasquilla}},\ and\ \bibinfo {author}
		{\bibfnamefont {L.}~\bibnamefont {Aolita}},\ }\bibfield  {title} {\bibinfo
		{title} {Quantum process tomography with unsupervised learning and tensor
			networks},\ }\href@noop {} {\bibfield  {journal} {\bibinfo  {journal} {Nat.
				Commun.}\ }\textbf {\bibinfo {volume} {14}},\ \bibinfo {pages} {2858}
		(\bibinfo {year} {2023})}\BibitemShut {NoStop}%
	\bibitem [{\citenamefont {Torlai}\ \emph {et~al.}(2018)\citenamefont {Torlai},
		\citenamefont {Mazzola}, \citenamefont {Carrasquilla}, \citenamefont
		{Troyer}, \citenamefont {Melko},\ and\ \citenamefont
		{Carleo}}]{torlai2018neural}%
	\BibitemOpen
	\bibfield  {author} {\bibinfo {author} {\bibfnamefont {G.}~\bibnamefont
			{Torlai}}, \bibinfo {author} {\bibfnamefont {G.}~\bibnamefont {Mazzola}},
		\bibinfo {author} {\bibfnamefont {J.}~\bibnamefont {Carrasquilla}}, \bibinfo
		{author} {\bibfnamefont {M.}~\bibnamefont {Troyer}}, \bibinfo {author}
		{\bibfnamefont {R.}~\bibnamefont {Melko}},\ and\ \bibinfo {author}
		{\bibfnamefont {G.}~\bibnamefont {Carleo}},\ }\bibfield  {title} {\bibinfo
		{title} {Neural-network quantum state tomography},\ }\href@noop {} {\bibfield
		{journal} {\bibinfo  {journal} {Nature Physics}\ }\textbf {\bibinfo {volume}
			{14}},\ \bibinfo {pages} {447} (\bibinfo {year} {2018})}\BibitemShut
	{NoStop}%
	\bibitem [{\citenamefont {Torlai}\ \emph {et~al.}(2019)\citenamefont {Torlai},
		\citenamefont {Timar}, \citenamefont {Van~Nieuwenburg}, \citenamefont
		{Levine}, \citenamefont {Omran}, \citenamefont {Keesling}, \citenamefont
		{Bernien}, \citenamefont {Greiner}, \citenamefont {Vuleti{\'c}},
		\citenamefont {Lukin} \emph {et~al.}}]{torlai2019integrating}%
	\BibitemOpen
	\bibfield  {author} {\bibinfo {author} {\bibfnamefont {G.}~\bibnamefont
			{Torlai}}, \bibinfo {author} {\bibfnamefont {B.}~\bibnamefont {Timar}},
		\bibinfo {author} {\bibfnamefont {E.~P.}\ \bibnamefont {Van~Nieuwenburg}},
		\bibinfo {author} {\bibfnamefont {H.}~\bibnamefont {Levine}}, \bibinfo
		{author} {\bibfnamefont {A.}~\bibnamefont {Omran}}, \bibinfo {author}
		{\bibfnamefont {A.}~\bibnamefont {Keesling}}, \bibinfo {author}
		{\bibfnamefont {H.}~\bibnamefont {Bernien}}, \bibinfo {author} {\bibfnamefont
			{M.}~\bibnamefont {Greiner}}, \bibinfo {author} {\bibfnamefont
			{V.}~\bibnamefont {Vuleti{\'c}}}, \bibinfo {author} {\bibfnamefont {M.~D.}\
			\bibnamefont {Lukin}}, \emph {et~al.},\ }\bibfield  {title} {\bibinfo {title}
		{Integrating neural networks with a quantum simulator for state
			reconstruction},\ }\href@noop {} {\bibfield  {journal} {\bibinfo  {journal}
			{Phys. Rev. Lett.}\ }\textbf {\bibinfo {volume} {123}},\ \bibinfo {pages}
		{230504} (\bibinfo {year} {2019})}\BibitemShut {NoStop}%
	\bibitem [{\citenamefont {Guedes~de Andrade}\ \emph {et~al.}(2022)\citenamefont
		{Guedes~de Andrade}, \citenamefont {D{\'\i}as}, \citenamefont {Navas},
		\citenamefont {Guha}, \citenamefont {Monta{\~n}o}, \citenamefont {Smith},
		\citenamefont {Raymer},\ and\ \citenamefont {Towsley}}]{de2022quantum}%
	\BibitemOpen
	\bibfield  {author} {\bibinfo {author} {\bibfnamefont {M.}~\bibnamefont
			{Guedes~de Andrade}}, \bibinfo {author} {\bibfnamefont {J.}~\bibnamefont
			{D{\'\i}as}}, \bibinfo {author} {\bibfnamefont {J.}~\bibnamefont {Navas}},
		\bibinfo {author} {\bibfnamefont {S.}~\bibnamefont {Guha}}, \bibinfo {author}
		{\bibfnamefont {I.}~\bibnamefont {Monta{\~n}o}}, \bibinfo {author}
		{\bibfnamefont {B.}~\bibnamefont {Smith}}, \bibinfo {author} {\bibfnamefont
			{M.}~\bibnamefont {Raymer}},\ and\ \bibinfo {author} {\bibfnamefont
			{D.}~\bibnamefont {Towsley}},\ }\bibfield  {title} {\bibinfo {title} {Quantum
			network tomography with multi-party state distribution},\ }\href@noop {}
	{\bibfield  {journal} {\bibinfo  {journal} {arXiv e-prints}\ ,\ \bibinfo
			{pages} {arXiv}} (\bibinfo {year} {2022})}\BibitemShut {NoStop}%
	\bibitem [{\citenamefont {Mucherino}\ \emph {et~al.}(2009)\citenamefont
		{Mucherino}, \citenamefont {Papajorgji},\ and\ \citenamefont
		{Pardalos}}]{mucherino2009k}%
	\BibitemOpen
	\bibfield  {author} {\bibinfo {author} {\bibfnamefont {A.}~\bibnamefont
			{Mucherino}}, \bibinfo {author} {\bibfnamefont {P.~J.}\ \bibnamefont
			{Papajorgji}},\ and\ \bibinfo {author} {\bibfnamefont {P.~M.}\ \bibnamefont
			{Pardalos}},\ }\bibfield  {title} {\bibinfo {title} {K-nearest neighbor
			classification},\ }in\ \href@noop {} {\emph {\bibinfo {booktitle} {Data
				mining in agriculture}}}\ (\bibinfo  {publisher} {Springer},\ \bibinfo {year}
	{2009})\BibitemShut {NoStop}%
	\bibitem [{\citenamefont {Mukherjee}\ \emph {et~al.}(2024)\citenamefont
		{Mukherjee}, \citenamefont {Schachenmayer}, \citenamefont {Whitlock},\ and\
		\citenamefont {W{\"u}ster}}]{mukherjeeNN:long}%
	\BibitemOpen
	\bibfield  {author} {\bibinfo {author} {\bibfnamefont {K.}~\bibnamefont
			{Mukherjee}}, \bibinfo {author} {\bibfnamefont {J.}~\bibnamefont
			{Schachenmayer}}, \bibinfo {author} {\bibfnamefont {S.}~\bibnamefont
			{Whitlock}},\ and\ \bibinfo {author} {\bibfnamefont {S.}~\bibnamefont
			{W{\"u}ster}},\ }\bibfield  {title} {\bibinfo {title} {Quantum network
			tomography of rydberg arrays by machine learning}} (\bibinfo {year} {2024}),\
	\bibinfo {note} {in preparation}\BibitemShut {NoStop}%
	\bibitem [{\citenamefont {Schuld}\ \emph {et~al.}(2014)\citenamefont {Schuld},
		\citenamefont {Sinayskiy},\ and\ \citenamefont
		{Petruccione}}]{schuld2014quest}%
	\BibitemOpen
	\bibfield  {author} {\bibinfo {author} {\bibfnamefont {M.}~\bibnamefont
			{Schuld}}, \bibinfo {author} {\bibfnamefont {I.}~\bibnamefont {Sinayskiy}},\
		and\ \bibinfo {author} {\bibfnamefont {F.}~\bibnamefont {Petruccione}},\
	}\bibfield  {title} {\bibinfo {title} {The quest for a quantum neural
			network},\ }\href@noop {} {\bibfield  {journal} {\bibinfo  {journal} {Quantum
				Inf. Process.}\ }\textbf {\bibinfo {volume} {13}},\ \bibinfo {pages} {2567}
		(\bibinfo {year} {2014})}\BibitemShut {NoStop}%
	\bibitem [{\citenamefont {Bhavna}\ and\ \citenamefont
		{Sonawane}(2023)}]{bhavna:NNridges}%
	\BibitemOpen
	\bibfield  {author} {\bibinfo {author} {\bibfnamefont {R.}~\bibnamefont
			{Bhavna}}\ and\ \bibinfo {author} {\bibfnamefont {M.}~\bibnamefont
			{Sonawane}},\ }\bibfield  {title} {\bibinfo {title} {A deep learning
			framework for quantitative analysis of actin microridges},\ }\href@noop {}
	{\bibfield  {journal} {\bibinfo  {journal} {npj Systems Biology and
				Applications}\ }\textbf {\bibinfo {volume} {9}},\ \bibinfo {pages} {21}
		(\bibinfo {year} {2023})}\BibitemShut {NoStop}%
	\bibitem [{\citenamefont {Peng}\ \emph {et~al.}(2015)\citenamefont {Peng},
		\citenamefont {Niu}, \citenamefont {Chen}, \citenamefont {Wu}, \citenamefont
		{Tung},\ and\ \citenamefont {Yang}}]{peng2015biological}%
	\BibitemOpen
	\bibfield  {author} {\bibinfo {author} {\bibfnamefont {H.-Q.}\ \bibnamefont
			{Peng}}, \bibinfo {author} {\bibfnamefont {L.-Y.}\ \bibnamefont {Niu}},
		\bibinfo {author} {\bibfnamefont {Y.-Z.}\ \bibnamefont {Chen}}, \bibinfo
		{author} {\bibfnamefont {L.-Z.}\ \bibnamefont {Wu}}, \bibinfo {author}
		{\bibfnamefont {C.-H.}\ \bibnamefont {Tung}},\ and\ \bibinfo {author}
		{\bibfnamefont {Q.-Z.}\ \bibnamefont {Yang}},\ }\bibfield  {title} {\bibinfo
		{title} {Biological applications of supramolecular assemblies designed for
			excitation energy transfer},\ }\href@noop {} {\bibfield  {journal} {\bibinfo
			{journal} {Chem. Rev.}\ }\textbf {\bibinfo {volume} {115}},\ \bibinfo {pages}
		{7502} (\bibinfo {year} {2015})}\BibitemShut {NoStop}%
	\bibitem [{\citenamefont {Schwartz}(2003)}]{schwartz2003conjugated}%
	\BibitemOpen
	\bibfield  {author} {\bibinfo {author} {\bibfnamefont {B.~J.}\ \bibnamefont
			{Schwartz}},\ }\bibfield  {title} {\bibinfo {title} {Conjugated polymers as
			molecular materials: How chain conformation and film morphology influence
			energy transfer and interchain interactions},\ }\href@noop {} {\bibfield
		{journal} {\bibinfo  {journal} {Annu. Rev. Phys. Chem.}\ }\textbf {\bibinfo
			{volume} {54}},\ \bibinfo {pages} {141} (\bibinfo {year} {2003})}\BibitemShut
	{NoStop}%
	\bibitem [{\citenamefont {Heightman}\ \emph {et~al.}(2024)\citenamefont
		{Heightman}, \citenamefont {Jiang},\ and\ \citenamefont
		{Acin}}]{Heightman_hamillearn_neuraldiffeq}%
	\BibitemOpen
	\bibfield  {author} {\bibinfo {author} {\bibfnamefont {T.}~\bibnamefont
			{Heightman}}, \bibinfo {author} {\bibfnamefont {E.}~\bibnamefont {Jiang}},\
		and\ \bibinfo {author} {\bibfnamefont {A.}~\bibnamefont {Acin}},\ }\bibfield
	{title} {\bibinfo {title} {Solving the quantum many-body hamiltonian learning
			problem with neural differential equations}} (\bibinfo {year} {2024}),\
	\bibinfo {note} {https://arxiv.org/abs/2408.08639}\BibitemShut {NoStop}%
	\bibitem [{\citenamefont {Carleo}\ and\ \citenamefont
		{Troyer}(2017)}]{carleo2017solving}%
	\BibitemOpen
	\bibfield  {author} {\bibinfo {author} {\bibfnamefont {G.}~\bibnamefont
			{Carleo}}\ and\ \bibinfo {author} {\bibfnamefont {M.}~\bibnamefont
			{Troyer}},\ }\bibfield  {title} {\bibinfo {title} {{Solving the quantum
				many-body problem with artificial neural networks}},\ }\href
	{https://doi.org/10.1126/science.aag2302} {\bibfield  {journal} {\bibinfo
			{journal} {Science}\ }\textbf {\bibinfo {volume} {355}},\ \bibinfo {pages}
		{602} (\bibinfo {year} {2017})}\BibitemShut {NoStop}%
	\bibitem [{\citenamefont {Bertalan}\ \emph {et~al.}(2019)\citenamefont
		{Bertalan}, \citenamefont {Dietrich}, \citenamefont {Mezi{\'c}},\ and\
		\citenamefont {Kevrekidis}}]{bertalan2019learning}%
	\BibitemOpen
	\bibfield  {author} {\bibinfo {author} {\bibfnamefont {T.}~\bibnamefont
			{Bertalan}}, \bibinfo {author} {\bibfnamefont {F.}~\bibnamefont {Dietrich}},
		\bibinfo {author} {\bibfnamefont {I.}~\bibnamefont {Mezi{\'c}}},\ and\
		\bibinfo {author} {\bibfnamefont {I.~G.}\ \bibnamefont {Kevrekidis}},\
	}\bibfield  {title} {\bibinfo {title} {On learning hamiltonian systems from
			data},\ }\href@noop {} {\bibfield  {journal} {\bibinfo  {journal} {Chaos: An
				Interdisciplinary Journal of Nonlinear Science}\ }\textbf {\bibinfo {volume}
			{29}} (\bibinfo {year} {2019})}\BibitemShut {NoStop}%
\end{thebibliography}
%
%

\end{document}